\documentclass[12pt]{article}
\usepackage[a4paper]{geometry}
\usepackage[english]{babel}

\usepackage{Alegreya}

\usepackage{geometry}
\usepackage{amsmath}
\usepackage{amsfonts}
\usepackage{amstext}
\usepackage{amssymb}
\usepackage{amsthm}
\usepackage{amscd}

\usepackage[pagebackref,draft=false]{hyperref}
\hypersetup{colorlinks,
linkcolor=myrefcolor,
citecolor=mycitecolor,
urlcolor=myurlcolor}

\usepackage[capitalize]{cleveref}
\usepackage{caption}

\usepackage{xcolor}
\definecolor{myurlcolor}{rgb}{0,0,0.4}
\definecolor{mycitecolor}{rgb}{0,0.5,0}
\definecolor{myrefcolor}{rgb}{0.5,0,0}
\usepackage{graphicx}
\usepackage{tikz}
\usepackage{tikz-cd}
\usepackage{mathrsfs}

\usepackage{etoolbox}
\usepackage{makeidx}
\usepackage{sectsty}
\usepackage{dsfont}
\usepackage{enumitem} 
\usepackage[]{latexsym}
\usepackage{braket}
\usepackage{caption}
\usepackage[utf8]{inputenx}
\usepackage[T1]{fontenc}
\usepackage{lmodern}
\usepackage{textcomp}
\usepackage{microtype}
\usepackage{totcount}
\usepackage{blindtext}

\newtheorem{rem}{Remark}

\newtheorem{prop}{Proposition}

\newtheorem{ex}{Example}

\newtheorem*{proof*}{Proof}

\newcommand{\be}{\begin{equation}}
\newcommand{\ee}{\end{equation}}
\newcommand{\bea}{\begin{eqnarray}}
\newcommand{\eea}{\end{eqnarray}}
\newcommand{\vsp}{\vspace{0.4cm}}
\newcommand{\grit}[1]{{\bfseries {\itshape {#1}}}}

\newcommand*{\Hrulefillsubpar}[1][0.4pt]{%
  \leavevmode\leaders\hrule height #1\hfill\kern0pt}
\newcommand*{\Hrulefillpar}[1][0.8pt]{%
  \leavevmode\leaders\hrule height #1\hfill\kern0pt}
\newcommand*{\Hrulefillsec}[1][1.6pt]{%
  \leavevmode\leaders\hrule height #1\hfill\kern0pt}
\newcommand*{\Hrulefillsubsec}[1][1.2pt]{%
  \leavevmode\leaders\hrule height #1\hfill\kern0pt}

\title{Evolutionary equations and constraints: Maxwell equations}

\author{F. M. Ciaglia$^{1,7}$  \href{https://orcid.org/0000-0002-8987-1181}{\includegraphics[scale=0.7]{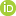}}, F. Di Cosmo$^{2,3,8}$ \href{https://orcid.org/0000-0002-8987-1181}{\includegraphics[scale=0.7]{ORCID.png}}, G. Marmo$^{4,5,9}$ \href{https://orcid.org/0000-0003-2662-2193}{\includegraphics[scale=0.7]{ORCID.png}}, L. Schiavone$^{6,10}$ \href{https://orcid.org/0000-0002-1817-5752}{\includegraphics[scale=0.7]{ORCID.png}}, \\ 
\footnotesize{$^{1}$\textit{ Max Planck Institute for Mathematics in the Sciences, Leipzig, Germany}} \\
\footnotesize{$^{2}$\textit{ Dep.to de Matematica, Univ. Carlos III de Madrid, Madrid, Spain}} \\
\footnotesize{$^{3}$\textit{ Instituto de Ciencias Matematicas (CSIC-UAM-UC3M-UCM) ICMAT, Madrid, Spain}} \\
\footnotesize{$^{4}$\textit{ INFN-Sezione di Napoli, Naples, Italy}} \\
\footnotesize{$^{5}$\textit{ Dipartimento di Fisica ``E. Pancini'', Universit\`a di Napoli Federico II,  Naples, Italy}} \\
\footnotesize{$^{6}$\textit{ Department of Mathematics, Faculty of Science, University of Ostrava, Ostrava, Czech Republic}} \\
\footnotesize{$^{7}$\textit{ e-mail: \texttt{ciaglia[at]mis.mpg.de}}} \\
\footnotesize{$^{8}$\textit{ e-mail: \texttt{f.cosmo[at]math.uc3m.es}}} \\
\footnotesize{$^{9}$\textit{ e-mail: \texttt{marmo[at]na.infn.it}}} \\
\footnotesize{$^{10}$\textit{ e-mail: \texttt{lucaschiavone[at]live.it}}} 
}

\begin{document}

\maketitle

\date{}

\begin{abstract}
By fixing a reference frame in spacetime, it is possible to split the Euler-Lagrange equations associated with a degenerate Lagrangian into purely evolutionary equations and constraints on the allowed Cauchy data with respect to the notion of Space and Time associated with the given reference frame.
In the context of Classical Electrodynamics, we introduce a ``covariantization procedure'' that allows to invert the perspective and to recover a full-fledged covariant formulation of Maxwell's equations starting only with constraint equations (i.e., Gauss' law and the local version of the law of conservation of magnetic flux) as perceived in a sufficient number of (inertial) reference frames on Minkowski spacetime.
\end{abstract}

\section{Introduction}

The splitting of the equations of motion of some given dynamical system into   a set of  genuine evolutionary equations  and a set of ``constraints'' is very common in Physics. 
For instance, in the framework of Lagrangian theories, this is always the case if we consider the Lagrangian theory associated with a degenerate Lagrangian. 
In this case, a careful analysis of the structure of the \textit{Euler-Lagrange equations} ``coming'' from a Lagrangian shows that part of the equations are first order ones, so that they only impose constraints on generalized ``positions'' and ``velocities''. 
Considering \textit{Classical Electrodynamics} as a study case, we show that it is possible to recover the full set of dynamical equations starting only from the constraint relations and by suitably exploiting the covariance properties of the field equations.
In particular, we recover the covariant formulation of \textit{Maxwell equations} from the \textit{Gauss law} and  the infinitesimal version of the law of conservation of magnetic flux.

\vsp

When describing \textit{Classical Electrodynamics}, there are two different, but related, points of view one may adopt: a covariant point of view, or a non-covariant point of view.
When adopting the latter, one starts by considering a three-dimensional Euclidean space $\mathbb{E}\,\cong\,(\mathbb{R}^{3},\delta)$, where $\delta$ is the standard Euclidean metric on $\mathbb{R}^{3}$, and four fields living on it, namely, the \textit{electric field}  $\textbf{E}$, the \textit{magnetic field} $\textbf{H}$, the \textit{electric induction} $\textbf{D}$ and the \textit{magnetic induction} $\textbf{B}$ (see for instance \cite{Sommerfeld, Jackson, LandauCampi}).  
Additionally, a constitutive law relating the fields  $(\textbf{E},\,\textbf{B})$ to the fields $(\textbf{D},\,\textbf{H})$ has to be provided in order to correctly describe experimental instances.
For instance, when dealing with electrodynamics in vacuum, it is usually assumed that $\textbf{E}\,\propto\,\textbf{D}$ and $\textbf{B}\,\propto\,\textbf{H}$. 
The dynamical content of the theory is given by the so-called \textit{Maxwell's equations}\footnote{In the following, we will always set $\epsilon = \mu = c = 1$ in vacuum.}:
\be
\begin{split}\label{eqn: original Maxwell}
\mathrm{div} \textbf{B} = 0 &\qquad \mathrm{div} \textbf{D} = \rho \\
\frac{d}{dt} \textbf{B} = - \mathrm{curl} \textbf{E} &\qquad \frac{d}{dt} \textbf{D} =  \mathrm{curl} \textbf{H} + \textbf{j}\,.
\end{split}
\ee 
From these equations, it is clear that the Euclidean metric tensor $\delta$ enters the game because the divergence and curl operators are defined in terms of it.

Note that the first two equations in \eqref{eqn: original Maxwell} do not involve time derivatives and, thus, they should be interpreted as constraints on the Cauchy data for the remaining equations.
This formulation of Maxwell equations is not manifestly covariant with respect to the Poincar\'{e} group.
Consequently, recalling that first-class constraints may be thought of as generators of gauge transformations (see \cite{DiracLectures}), one should follow the procedure outlined in \cite{BalachandranGaussLaw, LizziConstraints} in order to deal with a covariant description of the above-mentioned constraints.

\begin{rem}
From the experimental point of view, we should look at \textit{Maxwell's equations} as being the local version of the phenomenological integral equations  
\be \label{Eq: Maxwell integral}
\begin{split}
\int_S {\textbf{B} \cdot d\textbf{S}} = 0 \;\; \forall \, S \,& \text{closed and without boundary} \\
\frac{d}{dt} \int_S \textbf{B} \cdot d\textbf{S} &= - \int_{\partial S} \textbf{E} \cdot d\textbf{l} \\
\int_{\partial V} {\textbf{D} \cdot d\textbf{S}} &= q_{in} =  \int_V {\rho dV} \;\;  \\
\frac{d}{dt} \int_S \textbf{D} \cdot   d\textbf{S} - \int_{S} & \textbf{j} \cdot  d\textbf{S} =  \int_{\partial S} \textbf{H} \cdot d\textbf{l}\,.
\end{split}
\ee 
In the case where the integrands are smooth, as we shall assume throughout the rest of this paper, it is possible to associate smooth differential forms to them, that is, we may set
\be
\begin{split}
\textbf{B} \cdot d\textbf{S} &\,=:\, \mathrm{B} \qquad \textbf{D} \cdot d\textbf{S} \,=:\, \mathrm{D} \qquad \rho \mathrm{d}V \,=:\, \rho \\
\textbf{E} \cdot d\textbf{l} &\,=:\, \mathrm{E} \qquad \textbf{H} \cdot d\textbf{l} \,=:\, \mathrm{H} \qquad  \textbf{j}\cdot \mathrm{d}\textbf{S} \,=:\, \mathrm{j}.
\end{split}
\ee 
Of course, this situation does not allow the description of a source constisting of a point charge.
In this case, the description of the charge distribution $\rho$ would be possible only by means of Dirac $\delta$ distribution. 
Therefore, in order to take into account all the possible sources for the Electromagnetic fields, one should deal with a \textit{weak formulation} in terms of distributional differential forms (see for instance \cite{ParrotElectrodynamics, WheelerGravitation}).
In this work we decided to set aside this possibility in order to avoid technical difficulties that would distract from the ideas we want to convey, and we  postpone a careful analysis of the weak formulation to future works. 
We should also mention that the identification  of $\mathrm{H}$ and $\mathrm{D}$ with fair differential forms is not possible if  we consider also orientation reversing diffeomorphisms (for instance, time reversal and parity transformations).
A detailed discussion may be found in \cite{MarmoParasecoliTulczyjew, MarmoTulczyjew, TulczyjewDeNicola}.
\end{rem}

On the other hand, when adopting the covariant point of view, one exploits the insights of the \textit{Special Theory of Relativity} in order to formulate Electrodynamics in terms of  differential forms on Minkowski spacetime manifold $\mathcal{M}$.
This geometrical approach provides an elegant description of the foundational aspects of \textit{Classical Electrodynamics} and is of the utmost importance for the transition to the general relativistic case.
Specifically,  the main ingredients are two differential two-forms, say $\mathrm{F}$ and $\mathrm{G}$, called the \textit{Faraday} and the \textit{Ampere} form respectively. 
In terms of these differential forms, the dynamical content of the theory is given by the equations
\be
\mathrm{d F} = 0\,, \qquad \mathrm{d G} = \mathrm{J}
\ee
where $\mathrm{J} = \rho - \mathrm{d}t \wedge \mathrm{j}$ (see \cite{ParrotElectrodynamics, AuchmannKurz, Fecko, MarmoShanxi, Scheck, HehlObukhov}). 
From these two equations we can recover Maxwell's equations in a sense that will be made more precise in section \ref{sec: reference frames and electrodynamics}.
Furthermore,  in this formulation of the dynamics of the theory,  the distinction between genuine evolutionary equations  and constraints is evidently lost. 

Note that, at this stage, the geometrical description of \textit{Classical Electrodynamics} makes no reference to a metric tensor on spacetime and it is thus clear that we may consider \textit{Classical Electrodynamics} on the flat Minkowski spacetime as well as \textit{Classical Electrodynamics} on an arbitrary general relativistic spacetime without changing the abstract scheme recalled above.
The metric tensor comes into play only when we consider the so-called constitutive equations for $\mathrm{F}$ and $\mathrm{G}$, that is, when we want to express $\mathrm{G}$ as a function of $\mathrm{F}$.
In vacuum, it is often used the constitutive equation $\mathrm{G}=\star \mathrm{F}$ where $\star$ is the Hodge star operator on differential forms, which is a linear operator depending on the choice of a metric tensor.
This is an example of \grit{local} constitutive equation, that is, an equation relating the values of the field $\mathrm{F}$ at a spacetime point $m$ with the values of the field $\mathrm{G}$ at the same spacetime point, however, in principle, it would be possible to consider \grit{non-local} constitutive equations, that is, equations relating the values of the field $\mathrm{F}$ at a spacetime point $m$ with the values of the field $\mathrm{G}$ at other spacetime points.

Furthermore, similarly to what happens for $\mathrm{H}$ and $\mathrm{D}$ in the non-covariant case, the identification of $\mathrm{G}$ with a "even"\footnote{Using the terminology of \cite{MarmoTulczyjew, MarmoParasecoliTulczyjew, TulczyjewDeNicola}, it means differential forms that do not change sign under transformations that change the orientation.} differential form on spacetime is possible only if we exclude the possibility of considering spacetime transformations that do not preserve the orientation induced by the metric tensor.
If we want to consider such transformations, we must consider $\mathrm{G}$ to be a "odd" (or ``twisted'') differential form and the theory becomes more complex (see, againg, \cite{MarmoParasecoliTulczyjew, MarmoTulczyjew, TulczyjewDeNicola}).
Accordingly, in the following, we will restrict our attention to spacetime transformation that preserve the spacetime orientation.

\vsp

The connection between the non-covariant and the covariant point of view will be investigated in section \ref{sec: reference frames and electrodynamics} in terms of the notion of inertial reference frame on Minkowski spacetime.
The covariant formulation of \textit{Classical Electrodynamics} is well-known and extensively studied, as the list of references in this paper will testify, nonetheless, we think that, from a pedagogical point of view, it is important to recall this formulation in order to put our main result, given in section \ref{Sec: from constraints to covariant}, in the correct context. 
Specifically, in section \ref{sec: reference frames and electrodynamics} we will see how to recover (something which is directly related with) the fields $\mathrm{E}$, $\mathrm{B}$, $\mathrm{D}$ and $\mathrm{H}$ starting from the differential two-forms $\mathrm{F}$ and $\mathrm{G}$. 
As it will become clear, this recovery procedure depends on the particular reference frame we are considering, and, concerning the dynamical content of the theory, this leads to the fact that what is ``perceived'' as a constraint equation in one reference frame need not be ``perceived'' as a constraint equation in another reference frame.
Starting from this observation, we formulate the following question:

\vsp

Is it possible to recover the dynamical content of the covariant formulation of \textit{Classical Electrodynamics} by suitably manipulating the information contained in the constraint equations as perceived by a sufficient number of observers in a given class of reference frames?

\vsp
The answer we give is in the affirmative provided we consider a particular class of reference frames on Minkowski spacetime, namely, the  inertial reference frames.
From the mathematical point of view, we may rephrase our main result as follows. 
We start with a family\footnote{The set $A$ is an index set, and we will see that we may (judiciously) take it to be a discrete set with only $4$ elements.} $\{\mathcal{F}^{(\alpha)}\}_{\alpha\in A}$ of 3-dimensional, regular foliations of Minkowski spacetime  $\mathcal{M}$. 
For every $\alpha\in A$, we assume that the leaves of $\mathcal{F}^{(\alpha)}$ are diffeomorphic with $\mathbb{R}^{3}$, and there exists a 1-dimensional, regular foliation $\mathcal{T}^{(\alpha)}$ the leaves of which are diffeomorphic with $\mathbb{R}$, and are transversal to the leaves of $\mathcal{F}^{(\alpha)}$. 
In particular, this means that we may parametrize every leaf of $\mathcal{F}^{(\alpha)}$ by means of a real parameter, say $\tau$. 
These foliations will represent the splittings of spacetime into space and time associated with different reference frames and the family $\{\mathcal{F}^{(\alpha)}\}_{\alpha\in A}$ will be indexed by elements of the Poincaré group "connecting" each element in the family to a "fiducial" reference frame. 
Let $\mathcal{S}^{(\alpha)}_t$ denote the leaf of the foliation $\mathcal{F}^{(\alpha)}$ for the value of the parameter $\tau=t$. 
Then, for every $\alpha \in A$ we define the pullback $\mathrm{B}^{(\alpha)}_t$ and $\mathrm{D}^{(\alpha)}_t$ of the two-forms $F$ and $G$ defined over $\mathcal{M}$, via the canonical immersion of$ \mathcal{S}_t^{(\alpha)}$ into $\mathcal{M}$. 
We will show that  
\begin{eqnarray*}
&\mathrm{d}\,\mathrm{B}^{(\alpha)}_{t}\,=\,0\\
&\mathrm{d}\,\mathrm{D}^{(\alpha)}_{t}\,=\,\rho^{(\alpha)}_{t}\,,
\end{eqnarray*}
$\rho^{(\alpha)}_{t}$ being a differential $3$-form on $\mathcal{S}^{(\alpha)}_t$, for every leaf parametrized by $t$ in an open interval $I^{(\alpha)}\subset\mathbb{R}$ is equivalent to
\begin{eqnarray*}
&\mathrm{d}\,\mathrm{F}\,=\,0\\
&\mathrm{d}\,\mathrm{G}\,=\,\mathrm{J}\,,
\end{eqnarray*} 
where $\mathrm{J}$ is a differential $3$-form on $\mathcal{M}$ satisfying the compatibility conditions:
\be
(i^{(\alpha)}_{t})^{*}(\mathrm{J})\,=\,\rho^{(\alpha)}_{t}\,
\ee
for all $t$ in an open interval $I^{(\alpha)}\subset\mathbb{R}$ and for all $\alpha\in A$.

\vsp

After this short digression let us outline the structure of the paper. 
In section \ref{Sec: reference frames}, we recall the geometrical definition of \textit{reference frame} on a given Spacetime.
In section \ref{sec: reference frames and electrodynamics}, we exploit the notion of reference frame in order to obtain the original formulation of Maxwell's equations  from the covariant formulation of Electrodynamics (see also \cite{Fecko, AuchmannKurz, ParrotElectrodynamics, HehlObukhov}). 
In section \ref{Sec: from constraints to covariant}, we will prove our main result, namely, we introduce a ``covariantization procedure'' to recover a full-fledged covariant formulation of Electrodynamics starting from the constraint equations (i.e., Gauss' law and the infinitesimal version of the law of conservation of magnetic flux) as perceived by a sufficient number of inertial reference frames on Minkowski spacetime.

\section{Reference frames and spacetime splitting} \label{Sec: reference frames}

Our starting point will be a naked spacetime $\mathcal{M}$, that is, a $4$-dimensional, contractible, differential manifold.
Quite often, but not necessarily, $\mathcal{M}$ is equipped with a metric tensor $g$ of Lorentzian type, and the couple $(\mathcal{M},\,g)$  is obtained as a  solution of \textit{Einstein equations}.
For instance, the solution of Einstein equations in vacuum is the spacetime of the special theory of relativity, i.e., Minkowski spacetime.
In this case, the manifold $\mathcal{M}$ is diffeomorphic to $\mathbb{R}^{4}$ and the metric tensor $g$ coincides with the constant metric $\eta$ of signature $(1,-1,-1,-1)$ or $(-1,1,1,1)$ \footnote{Usually, the former one is used in Particle Physics, while the latter in General Relativity.}.

Once we have a naked spacetime $\mathcal{M}$, we have a $4$-dimensional differential manifold in which the separate notions of space and time are absent.
For the identification of physical observables, however, we do need a splitting into time and space.
After all, every experiment is performed in some definite place at some definite time.
The need of this distinction between space and time is clearly expressed by H. Weyl who writes\footnote{H. Weyl. Raum, Zeit, Materie. Springer, Berlin (1923). English translation by H. L. Brose (Space, time and matter. Dover, New York (1952)).}:

\begin{quotation}
``Time is the primitive form of the stream of consciousness. It is a fact, however
obscure and perplexing to our minds, that the contents of consciousness do not
present themselves simply as being, but as being now filling the form of the
enduring present with a varying content.
So that one does not say this is but this is now, yet now no more. 
If we project ourselves outside the stream of consciousness and represent its contents as an object, it becomes an event happening in time, the separate stages of which stand to one another in the relation of earlier and later.
Just as time is the form of the stream of consciousness, so one may justifiably
assert that space is the form of external material reality.''
\end{quotation}

It is then natural to ask how to describe a splitting of a given naked spacetime $\mathcal{M}$ into space and time.
The answer to this question may be found by means of a \textit{nowhere vanishing}, smooth (1,1) tensor field idempotent and of rank-one, say $\mathcal{R}$, such that 
\be\label{eqn: reference frame condition}
\mathcal{R}\circ\mathcal{R}=\mathcal{R},
\ee
where $\circ$ denotes the contraction among tensor fields, which defines a structure of a real, associative algebra  on $(1,1)$ tensor fields.
We call any such $(1,1)$ tensor field a \textit{reference frame} on $\mathcal{M}$ (see \cite{deritis_marmo_preziosi-a_new_look_at_relativity_transformations, MarmoPreziosi}).
In the following, we will decompose the tensor field $\mathcal{R}$ as 
\be
\mathcal{R}=\Gamma\otimes\theta,
\ee
where $\Gamma$ is a \textit{nowhere vanishing} vector field and $\theta$ is a \textit{nowhere vanishing} one-form on $\mathcal{M}$.
Then, the condition in equation \eqref{eqn: reference frame condition} is equivalent to 
\be
\theta(\Gamma)=1.
\ee
If $\mathcal{M}$ is endowed with a metric tensor $g$ of Lorentzian type, then $\Gamma$ is required to be a timelike vector field and it is customary to set $\theta(X):=g(\Gamma,X)$ for every vector field $X$ on $\mathcal{M}$ \cite{AuchmannKurz, Fecko, DeFeliceBini}.
However, we want to stress that, in principle, a reference frame $\mathcal{R}$ on $\mathcal{M}$ may be defined independently of any metric tensor $g$.

\begin{rem}\label{rem: reparametrization}
Note that if we set $\Gamma'=\mathrm{e}^{f}\Gamma$ with $f$ a smooth function on $\mathcal{M}$, and define $\theta'=\mathrm{e}^{-f}\theta$, we are not changing reference frame because $\mathcal{R}=\Gamma\otimes\theta=\Gamma'\otimes\theta'$.
The transformation $\Gamma\mapsto\Gamma'$ is a reparametrization of the integral curves of $\Gamma$ and, according to what will be said below, amounts to a change in the parametrization of the ``time'' associated with the reference frame $\mathcal{R}=\Gamma\otimes\theta$.
\end{rem}

Once we have the reference frame  $\mathcal{R}=\Gamma\otimes\theta$ we may decompose the tangent space $\textbf{T}_{m}\mathcal{M}$ at each $m\in\mathcal{M}$ into the direct sum:
\begin{equation}
\textbf{T}_{m}\mathcal{M}\,=\,\mathrm{Im}(\mathcal{R}_{m})\,\oplus\,\mathrm{Ker}(\mathcal{R}_{m})\,=\,\mathrm{span}(\Gamma_{m})\,\oplus\,\mathrm{Ker}(\theta_{m})\,.
\end{equation}
A moment of reflection shows that this decomposition of $\textbf{T}_{m}\mathcal{M}$ depends only on $\mathcal{R}$ and not on $\Gamma$ and $\theta$ (see remark \ref{rem: reparametrization}).
In accordance with remark \ref{rem: reparametrization}, we choose a factorization of $\mathcal{R}$ in terms of a complete vector field $\Gamma$, and we assume $\mathrm{Ker}(\theta)$ to have fixed rank and to be an integrable distribution, that is, we assume 
\be
\theta\wedge\mathrm{d}\theta=0.
\ee
This allows us to decompose the module $\mathfrak{X}(\mathcal{M})$ of vector fields on $\mathcal{M}$ into a direct sum:
\begin{equation}
\mathfrak{X}(\mathcal{M})\,=\,\mathrm{span}(\Gamma)\,\oplus\,\mathrm{Ker}(\theta)\,.
\end{equation}
The integrability of $\mathrm{Ker}(\theta)$ implies there is a codimension-one foliation $\mathcal{F}$ of $\mathcal{M}$, and this foliation is transversal to the one-dimensional foliation of $\mathcal{M}$ given by the integral curves of $\Gamma$ because $\theta(\Gamma)=1$ (see \cite[p. $42$]{CandelConlonI}).

\begin{rem}
In the following, we will always assume that both $\Gamma$ and $\mathrm{Ker}(\theta)$ give rise to regular foliations of $\mathcal{M}$ such that their leaves are smooth embedded submanifolds.
Furthermore, we will assume that all the leaves of $\mathcal{F}$  are diffeomorphic to each other, and that the vector field $\Gamma$ is the infinitesimal generator of  a non-trivial action of $\mathbb{R}$ on $\mathcal{M}$.
In general relativity, there is an important class of spacetimes for which we may find reference frames satisfying these conditions, namely, the so-called  globally hyperbolic spacetimes (see \cite{Hawking}).
Moreover, when $\mathrm{Ker}(\theta)$ is generated (as a module) by a family of complete vector fields closing on a Lie algebra, we obtain  Bianchi's classification (see \cite{MarmoCapozziello}).
\end{rem}

The integral curves of $\Gamma$ are called \textit{clocks} and define the ``time'' associated with the reference frame $\mathcal{R}=\Gamma\otimes\theta$, while every leaf of $\mathcal{F}$ is called a \textit{simultaneity surface} and is interpreted as the instantaneous physical space associated with the reference frame $\mathcal{R}=\Gamma\otimes\theta$.
If $\theta$ is exact, i.e., there exists a smooth function $\mathscr{T}$ such that  
\be
\theta= \mathrm{d}\mathscr{T},
\ee
the leaves of $\mathcal{F}$ are the level sets of $\mathscr{T}$, and thus $\mathscr{T}$ is called a \textit{time function}. 

In general, a $(1,1)$ tensor field may induce a foliation of the spacetime into space and time but only in some cases the manifold $\mathcal{M}$ can be given a global "product structure", that is:
\begin{equation}\label{eq: double fibration}
\begin{tikzcd}
\mathcal{M}\,\cong\,\mathcal{T}\times\mathcal{S} \arrow[r, "\pi_3"] \arrow[d, "\pi_1"] & \mathcal{S} \\
\mathcal{T}&\,
\end{tikzcd}
\end{equation}
in such a way that the vector space isomorphisms 
\be
\textbf{T}_{\pi_{1}(m)}\mathcal{T}\cong\mathrm{span}(\Gamma_{m})
\ee
and 
\be
\textbf{T}_{\pi_{3}(m)}\mathcal{S}\cong\mathrm{Ker}(\theta_{m})
\ee
hold for every $m\in\mathcal{M}$.
In this case, we say that the decomposition 
\be
\mathcal{M}\,\cong\,\mathcal{T}\times\mathcal{S}
\ee
is compatible with the reference frame $\mathcal{R}$.
Then, the splitting 
\be
\textbf{T}_m \mathcal{M}\cong \mathrm{span}(\Gamma_m)\,\oplus\,\mathrm{Ker}(\theta_m)
\ee
can be associated with a \textit{connection} on $\mathcal{M}$, but there are two ways of looking at it: either we consider $\mathrm{span}(\Gamma)$ as the vertical distribution and $\mathrm{Ker}(\theta)$ as the horizontal one, or we consider  $\mathrm{Ker}(\theta)$ as the vertical distribution and $\mathrm{span}(\Gamma)$ as an horizontal one.
In the first case we are thinking of $\mathcal{M}$ as a fiber bundle over the time manifold $\mathcal{T}$, while in the second case we are thinking of $\mathcal{M}$ as a fiber bundle over the space manifold $\mathcal{S}$. 
Once again, we want to stress that these constructions do not depend on the existence of a metric tensor on $\mathcal{M}$.

\begin{ex}\label{ex: inertial reference frame on minkowski spacetime}
To be more concrete, let us consider Minkowski spacetime $(\mathcal{M},\,g)\cong(\mathbb{R}^4,\,\eta)$ together with a set of Cartesian coordinates $(x^{0}, \textbf{x})$. 
We take the reference frame 
\be
\mathcal{R}_{0} = \Gamma_{0}\otimes\theta_{0}= \partial_0 \otimes \mathrm{d}x^0\,,
\ee
which is characterized by the time distribution 
\be
\mathrm{span}(\Gamma_{0}) = \mathrm{span}(\partial_0)
\ee
and the space distribution 
\be
\mathrm{Ker}(\theta_{0}) =\mathrm{span}\left( \partial_1, \partial_2, \partial_3 \right)\,.
\ee
In this case, we have the \textit{time function} $\mathscr{T}=x^0$, the simultaneity leaves are all diffeomorphic to $\mathbb{R}^3$ and 
\be
\mathcal{M}\cong \mathcal{T} \times \mathcal{S} \,\cong\, \mathbb{R} \times \mathbb{R}^3.
\ee
This reference frame $\mathcal{R}_0$ is called \textit{geodesical} (Lorenzian in our case, since $g$ has a Lorenzian signature) because 
\be
\nabla_{\Gamma_{0}}\Gamma_{0}=0
\ee
where $\nabla$ denotes the covariant derivative associated with the Levi-Civita connection on $(\mathcal{M},\,g)$, i.e., the integral curves of $\,\Gamma_{0}$ are timelike geodesic curves for the metric tensor $\eta$.
\end{ex}

\begin{rem}
In the case in which there is no metric tensor on $\mathcal{M}$, we may define an inertial reference frame if we consider as starting point a second order vector field on $T\mathcal{M}$, say $\Gamma$, and consider the connection on $\mathcal{M}$ associated with it (see \cite{MarmoTangent}).  
If $\mathcal{M}$ is diffeomorphic to $\mathbb{R}^{4}$, $\Gamma$ is complete and its associated connection is flat, the projection on $\mathcal{M}$ of the integral curves of the vector field $\Gamma$ describe a ``free system on $\mathcal{M}$'' (congruence of inertial motions).
Furthermore, we can find $\mathrm{dim}(\mathcal{M})=4$ global constants of the motion for $\Gamma$ and we may decompose $\Gamma$ in terms of $4$ vector fields $V_{0},\,V_{1},\,V_{2},\,V_{3}$ that are tangent to the hypersurfaces in $T\mathcal{M}$ selected by fixing the values of the constants of the motion, i.e.
\be
\Gamma = \sum_{\mu = 0}^3\Gamma_{\mu}V^{\mu}\,.
\ee
It turns out that these hypersurfaces are diffeomorphic with $\mathcal{M}$ so that  the vector field $V_{j}$ may be ``projected'' on $\mathcal{M}$ for every $j=0,...,3$ (\cite{Carinena_Gracia_Marmo_Martinez_Munoz_RomanRoy_Geometric_HamiltonJacobi_theory}).
Clearly, the projection procedure depends on the explicit choice of the values of the constants of the motion.
The integral curves of the projection  $\tilde{\Gamma}$ of $\Gamma$ on $\mathcal{M}$ are a congruence of inertial motions on $\mathcal{M}$ and their associated vector field may serve as the contravariant part of a reference frame once we find a suitable one-form $\theta$ such that $\theta(\tilde{\Gamma})=1$ (see \cite{deritis_marmo_preziosi-a_new_look_at_relativity_transformations, MarmoPreziosi}).
The reference frame $\mathcal{R}=\tilde{\Gamma}\otimes\theta$ is called \textit{inertial}.

Let us stress that this construction does not require any metric tensor and, indeed, it is compatible with all possible flat metric tensors on $\mathcal{M}\cong\mathbb{R}^{4}$ with different signatures.
\end{rem}

Given a reference frame $\mathcal{R}=\Gamma\otimes \theta$, it is possible to build a differential calculus adapted to the foliation associated with the reference frame. The kernel of the form $\theta$ can be always given the structure of a left module over the ring $\mathcal{F}(\mathcal{M})$ of smooth functions on $\mathcal{M}$. Let us assume that this module is free and that the elements of a basis, namely $\left\lbrace X_{j} \right\rbrace$ with $j=1,2,3$, close on a Lie algebra over the reals, with $\left[ \cdot, \cdot \right]$ denoting the Lie bracket.
An explicit example to keep in mind is the case of Minkowski spacetime of Example \ref{ex: inertial reference frame on minkowski spacetime}, where the elements of a basis close on an Abelian algebra, defining translational symmetry of space.
 
Let $\Omega_0^1(\mathcal{M})$ be the space of all linear functionals over $\mathrm{Ker} (\theta)$.
This set can be endowed with the structure of a $\mathcal{F}(\mathcal{M})$-bimodule,  and it is also a free module.
A possible dual basis can be chosen by means of the metric tensor, i.e., $\alpha^j (\cdot) = g(X_j, \cdot)$. 
Using these two basic building blocks, we can obtain the space of all $(m,n)$-tensor fields which are $m$-times contravariant and $n$-times covariant, say $\tau_0^{m,n}(\mathcal{M})$. 
A generic element 
$$
T\colon\underbrace{\Omega_0^1(\mathcal{M})\times \Omega_0^1(\mathcal{M}) \times \cdots}_\mathrm{m-times} \times \underbrace{\mathrm{Ker} (\theta) \times \mathrm{Ker}(\theta) \times \cdots}_\mathrm{n-times} \rightarrow \mathbb{R}
$$
is a real-valued multilinear function acting on m copies of the space $\Omega_0^1(\mathcal{M})$ and $n$ copies of $\mathrm{Ker}(\theta)$. 
In particular, $p$-contravariant skew-symmetric tensors ``live'' in the left $\mathcal{F}(\mathcal{M})$-module denoted by $\Lambda_0^{p}(\mathcal{M})$ and will be called $p$-multivectors, whereas $p$-covariant skew-symmetric tensors will be named $p$-forms and the bimodule of all such tensors is denoted by $\Omega^{p}_0(\mathcal{M})$.     

The two direct sums 
\begin{equation}
\Lambda_0(\mathcal{M}) = \oplus_{p=0}^3\Lambda_0^p(\mathcal{M}) \qquad  \Omega_0(\mathcal{M}) = \oplus_{p=0}^3\Omega_0^p(\mathcal{M}) 
\end{equation}
can be equipped with the structure of a graded algebra with respect to the wedge product.
In the case of multivectors, the product is defined by
\begin{equation}
\begin{split}
&X \wedge Y (\omega_1, \omega_2,\cdots , \omega_{k+l}) = \\ 
=& \sum_{\sigma} \mathrm{sign}(\sigma) X(\omega_{\sigma(1)}, \cdots, \omega_{\sigma(k)})Y(\omega_{\sigma(k+1),\cdots , \sigma(k+l)})
\end{split}
\end{equation}
where $X\in \Lambda^{k}_0(\mathcal{M})$, $Y\in \Lambda^l_0(\mathcal{M})$ and $\omega_j \in \Omega_0^1(\mathcal{M})$. 
Analogously, the wedge product  on the space of differential forms is defined by replacing forms and vectors in the previous equation. 
It is possible to define also a dual coupling $\langle \cdot, \cdot\rangle$ between a p-form and a p-multivector. 
Indeed, if $\Lambda_0^2(\mathcal{M}) \ni Y = Y^1 \wedge Y^2$ with $Y^j \in \Lambda^1_0(\mathcal{M})$, and $\Omega_0^2(\mathcal{M}) \ni \omega = \omega^1 \wedge \omega^2 $, with $\omega^j \in \Omega_0^1(\mathcal{M})$, the aforementioned coupling has the following expression 
\begin{equation}
\langle \omega , Y \rangle = |\det(\omega^j(Y^k))| 
\label{dual coupling}
\end{equation}
and this product is extended to higher rank objects. 

Let us now introduce the so called boundary operator 
\be
\partial^{(p+1)} : \Lambda^{p+1}_0(\mathcal{M}) \rightarrow \Lambda_0^p(\mathcal{M}) .
\ee
It is a linear nihilpotent operator acting on a multivector $Y = Y^1\wedge \cdots\wedge Y^{p+1}$ as follows:
\begin{equation}
\begin{split}
\partial Y = \sum_{j,k=1}^{p+1} (-1)^{j+k+1} \left[ Y^j, Y^k \right]&\wedge Y^1 \wedge \cdots \wedge \hat{Y}^j \wedge \cdots \\
&\cdots \wedge \hat{Y}^k \wedge \cdots \wedge Y^{p+1}\,,
\label{boundary operator}
\end{split}
\end{equation}
where the symbol $\hat{Y}^j$ means that the vector $Y^j$ is missing in the wedge product. 
It can be extended to a single operator 
\be
\partial : \Lambda_0(\mathcal{M}) \rightarrow \Lambda_0(\mathcal{M}) .
\ee
on the whole algebra of multivectors. 
Since $\partial^2 = 0$, this operator can be used to define homology groups 
\be
\mathbb{H}^p_0(\mathcal{M}) = \frac{\mathrm{Ker}(\partial^{(p)})}{\mathrm{Im}(\partial^{(p+1)})}\,,
\ee
but it is not a derivation of the wedge product.

By means of the dual coupling given in equation \eqref{dual coupling}, it is possible to extend this construction to the space of differential forms. 
In particular, since vector fields act on the ring of smooth functions $\mathcal{F}(\mathcal{M})$ via Lie derivative, it is possible to define a coboundary operator 
\be
d_{\perp}^{(p)}: \Omega_0^{p} \rightarrow \Omega^{p+1}_0
\ee
as follows:
\be
\begin{split}
d^{(p)}_{\perp} \omega (&Y^1,\, \cdots,\,Y^{p+1})= \sum_{j=1}^{p+1} (-1)^{j+1} L_{Y^j}(\omega (Y^1, \, \cdots \\
&\cdots , \hat{Y}^j, \cdots, \, Y^{p+1})) +(-1)^{j+k} \omega([Y^j, Y^k], \, Y^1,\\ 
&\cdots , \hat{Y}^j, \cdots \hat{Y}^k, \cdots, \, Y^{p+1})\,,\label{coboundary operator}
\end{split}
\ee
where $\omega \in \Omega^p_0(\mathcal{M})$ and the $\hat{Y}^j$ means again that the vector $Y^j$ is missing. 
These operators can be extended to the direct sum of all differential forms in order to  introduce a single coboundary operator 
\be
d_{\perp} : \Omega_0 \rightarrow \Omega_0 .
\ee
It can be proved that this operator is nilpotent, i.e., $d_{\perp}^2 = 0$, and the cohomology groups are defined as 
\be
H^{p}_0(\mathcal{M}) = \frac{\mathrm{Ker}(d^{(p)})}{\mathrm{Im}(d^{(p-1)})}
\ee
Contrarily to the boundary operator, $d_{\perp}$ is a graded derivation of degree 1 with respect to the wedge product, i.e., we have
\be
d_{\perp} (\alpha \wedge \beta) = d_{\perp}\alpha \wedge \beta + (-1)^{1\cdot |\alpha|} \alpha \wedge d_{\perp}\beta .
\ee
The module $\Omega_0(\mathcal{M})$ is a submodule of the module $\Omega(\mathcal{M})$ of differential forms over $\mathcal{M}$, also called exterior algebra. 
By means of the idempotent $(1-1)$ tensor field $\mathcal{P}= \mathbf{1} - \mathcal{R}$ associated with a reference frame, one can extend the operator $d_{\perp}$ to an operator over the whole exterior algebra $\Omega(\mathcal{M})$. 
With an evident abuse of notation this transversal coboundary operator will be denoted by the same symbol $d_{\perp}$, and it acts on a p-form $\omega \in \Omega^p(\mathcal{M})$ as follows: 
\be
\begin{split}
d_{\perp}\omega (X^1,\cdots,\,X^{p+1}) &= \left( \mathcal{P}^{*}\right )(d\omega (X^1,\cdots,\,X^{p+1})) = \\ 
&= d\omega \left(\mathcal{P}\left(X^1 \right),\cdots,\,\mathcal{P}\left(X^{p+1}\right)\right)\,.
\end{split}
\ee
In particular, when the reference frame $\mathcal{R}=\Gamma\otimes \theta$  is such that $d\theta = 0$, the following decomposition is valid
\begin{equation}
\mathrm{d}\omega = \theta \wedge i_{\Gamma}\mathrm{d}\omega + d_{\perp}\omega\,,\label{Eq: differential of omega}
\end{equation} 
and we obtain a decomposition  in terms of a "transversal" and a "temporal" component.
The transversal component acts as 
\be\label{Eq: decomposition domega perp}
\begin{split}
 d_{\perp}\omega(X^1,\cdots, X^{p+1}) & = d\omega \left(\mathcal{P}\left(X^1 \right),\cdots,\,\mathcal{P}\left(X^{p+1}\right)\right) = \\
& = i_{\mathcal{P}(X^{p+1})}\cdots i_{\mathcal{P}(X^{1})} i_{\Gamma}\left( \theta \wedge d\omega \right) = \\
& = i_{\Gamma}\left( \theta\wedge d\omega \right) \left( X^1, \cdots, X^{p+1} \right)\,, 
\end{split}
\ee
which means
\be
d_{\perp}(\omega) = i_{\Gamma}\left(\theta \wedge d\omega \right) .
\ee
It is important to stress that the above chain of equalities is true because $i_{\Gamma}i_{\Gamma}\alpha = 0$, $\forall \alpha \in \Omega\left(\mathcal{M}\right)$ and the reference frame satisfies $\theta(\Gamma)=1$. 

The previous result suggests to decompose any differential p-form $\omega \in \Omega^{p}(\mathcal{M})$ as the sum of a transversal component $\omega_{\perp}$ and temporal component $\omega_{\parallel}$ setting
\begin{equation}
\label{Eq:DecompositionDifferentialForms}
\omega = \omega_{\perp} + \omega_{\parallel}\,,
\end{equation}
where
\begin{eqnarray}
& \omega_{\perp} = (\mathcal{P})^*(\omega) = i_{\Gamma}\left( \theta \wedge \omega \right) \label{Eq:PerpComponent}\\
& \omega_{\parallel} = \omega - \omega_{\perp} = \theta \wedge i_{\Gamma}\omega\,.
\end{eqnarray}
Then, the expression in equation \eqref{Eq: differential of omega}  can be written as 
\be \label{Eq:DecompositionDifferentialOfDifferentialForms}
\mathrm{d} \omega = \theta \wedge \underbrace{( \mathcal{L}_\Gamma \omega_{\perp} - \mathrm{d}_{\perp} i_\Gamma \omega )}_{=(\mathrm{d}\omega)_\parallel} + \underbrace{\mathrm{d}\theta \wedge \omega_\parallel + \mathrm{d}_{\perp} \omega_{\perp}}_{=(\mathrm{d}\omega)_\perp} .
\ee
This way of decomposing tensor fields by means of a reference frame has received particular attention especially in connection with Electrodynamics \cite{AuchmannKurz, Fecko, DeFeliceBini} and we will make use of it in the next section. 
In the context of Special Relativity, we will exclusively deal with reference frames of the type we saw in example \ref{ex: inertial reference frame on minkowski spacetime} and with all those that may be obtained from it by means of a Poincaré transformation. 
In this case, the vector fields tangent to a leaf form an Abelian algebra and the boundary operator acts trivially. 
The reference frame of example \ref{ex: inertial reference frame on minkowski spacetime} satisfies the condition $\mathrm{d}\theta \,=\, 0$ and, as it will be clear in the end of the section, this condition is preserved  by Poincaré transformations. 
Therefore, we will be entitiled to use a decomposition of the type given in equation  \eqref{Eq:DecompositionDifferentialOfDifferentialForms} where $(\mathrm{d}\omega)_\perp \,=\, \mathrm{d}_\perp \omega_\perp$.

\vsp

Now, consider a naked spacetime $\mathcal{M}$ and a reference frame $\mathcal{R}$ on it $\mathcal{M}$, and assume that $\mathcal{M}$ admits a decomposition  into the Cartesian product of the time manifold $\mathcal{T}\cong\mathbb{R}$ with the space manifold $\mathcal{S}$ which is compatible with the reference frame $\mathcal{R}$ as explained above.
Then, for every $t\in \mathcal{T}$, there is a canonical immersion of the space manifold $\mathcal{S}$ into $\mathcal{M}$:
\be \label{Eq: canonical immersion}
i^{(\mathcal{R})}_t \;\; \vert \;\; \mathcal{S} \hookrightarrow \mathcal{M}\;\; : \;\; \textbf{x} \mapsto i^{(\mathcal{R})}_t(\textbf{x})=(\textbf{x}, \,t)\,.
\ee 
This  map clearly depends on the particular reference frame considered and on the  explicit value of $t\in\mathcal{T}$.
Essentially, $i^{(\mathcal{R})}_t$ is a section for the projection $\pi_3$ of equation \eqref{eq: double fibration} and the image of $\mathcal{S}$ through $i^{(\mathcal{R})}_t$ is a simultaneity surface in $\mathcal{M}$\footnote{A similar construction may be found in \cite{MarsdenElasticity}.}.
Thanks to this map, it is possible to define the pull-back $\omega_{\mathcal{R}}^{t}$ on the space manifold $\mathcal{S}$ of a differential form $\omega$  on $\mathcal{M}$.
This operation gives a differential form on $\mathcal{S}$ which is parametrized by the value of $t$. 
Since the tangent space at each point in $\mathcal{S}$ is spanned by vectors which are $i^{(\mathcal{R})}_t$-related with vectors in the kernel of $\theta$, the pull-back operation cancels out all terms involving $\theta$. 
An important property of the pull-back is that it "commutes" with the exterior derivative $\mathrm{d}$:
\be
{i^{(\mathcal{R})}_t}^* \circ \mathrm{d} = \mathrm{d} \circ {i^{(\mathcal{R})}_t}^*
\ee
where, on the r.h.s., $\mathrm{d}$ is the exterior derivative on $\mathcal{S}$. 
From this, it follows that if a differential $k-$form  $\omega$ on $\mathcal{M}$ is closed, then, the pullback  $\omega_{\mathcal{R}}^{t}$ is closed, or equivalently,
\be
\mathrm{d} \omega = 0 \implies \mathrm{d}\omega_{\mathcal{R}}^{t} = 0\,.
\ee
In general,  the converse is not true.
However, since the map $i^{(\mathcal{R})}_t$  depends on the reference frame $\mathcal{R}$, it is legitimate to think that, by considering a suitable family of reference frames $\left\{\mathcal{R}^a \right\}_{a \in \mathcal{A}}$, it would be possible that the condition $\mathrm{d}\omega_{\mathcal{R}^{a}}^{t^{a}} = 0 \;\; \forall a \in \mathcal{A}$ implies $\mathrm{d}\omega = 0$. 
Roughly speaking, as far as differential forms are considered, the  amount of information extracted from a sufficiently large set of simultaneity  surfaces is equivalent to the amount of information extracted from the whole spacetime.
In the next section, we will show how this is connected with the possibility of obtaining the covariant formulation of Maxwell equations out of the constraint equations of $3$-dimensional Electrodynamics on a sufficiently large set of simultaneity surfaces.

It is worth stressing that the canonical immersion \eqref{Eq: canonical immersion} is strictly related with the differential calculus adapted to the foliation we just developed. The leaves that we immerse into $\mathcal{M}$ via \eqref{Eq: canonical immersion} are integral submanifolds for the involutive distribution $\mathrm{ker}\theta$. This implies that
\be
(\,i^{(\mathcal{R})}_t \,)^* \theta \,=\, 0 \,.
\ee
Therefore, taking the pull-back of a differential form via this immersion amounts exactly to taking the pull-back of its transversal component and we have
\be
\begin{split}
\mathrm{d}\omega^t_{\mathcal{R}} \,=\, 0\, \implies \, \mathrm{d} (\, i^t_{\mathcal{R}}\,)^* \omega &\,=\, (\, i^t_{\mathcal{R}}\,)^* \mathrm{d} \omega \,=\, (\, i^t_{\mathcal{R}}\,)^* (\mathrm{d}\omega)_{\perp}\,= \\ 
&=\, (\, i^t_{\mathcal{R}}\,)^* \mathrm{d}_\perp \omega_\perp \,=\, 0 \,.
\end{split}
\ee

\vsp

We conclude this section with some comments about the choice of the family $\left\{\mathcal{R}^a \right\}_{a \in \mathcal{A}}$ of reference frames we will consider in the proof of our main result in section \ref{Sec: from constraints to covariant}
.
First of all, let us fix a fiducial reference frame, say $\mathcal{R}_0$.
From the physical point of view, we want to consider only those reference frames that give rise to splittings into time and space which are pairwise compatible.
Given a reference frame $\mathcal{R}_1$ we say that it is compatible with $\mathcal{R}_0$ if the so-called \textit{mutual objective existence conditions} (see \cite{MarmoPreziosi}) are satisfied:
\be
\begin{split}
\theta_0 \left( \,\Gamma_1\, \right)\,&\neq\,0\\
\theta_1 \left( \,\Gamma_0\, \right)\,&\neq\,0\,.
\end{split}
\ee
From the physical point of view, the mutual objective existence conditions impose that every world line of a reference frame does not lie in some simultaneity surface of the other reference frame, that is, it is not possible that the obervers associated with a given reference frame  exist  only for one particular value of time  (and never in the past and never again in the future) in the other reference frame.
It is then natural to search for a group $G$ which will define an orbit of pairwise compatible frames starting with a fiducial one.
This circumstance will guarantee that the compatibility notion we have introduced will be an equivalence relation.
Specifically, we look for a family 
\be
\{\mathcal{R}_a = \theta_a \otimes \Gamma_a\}_{a\in\mathcal{A}}
\ee
such that  
\be
\forall \, a \in \mathcal{A} \;\; \exists \, g_a \in G \;: \; \theta_a = \phi_{g_a}^*\theta_0
\ee
and 
\be
\Gamma_a = {\phi_{g_a}}_* \, \Gamma_0
\ee
where $\phi_g$ is the representation of the element $g_{a} \in G$ by means of a diffeomorphism on $\mathcal{M}$. 
In \cite[p. 598-600]{MarmoPreziosi}   it is shown how to recover the (connected component of the) Poincar\'{e} group as the symmetry group implementing the mutual objective existence conditions between the family of reference frames on $\mathcal{M}\cong\mathbb{R}^{4}$ of the form given in the example \ref{ex: inertial reference frame on minkowski spacetime}. 
Indeed, we will build on this construction in order to choose a family of mutually compatible reference frames interconnected via a finite number of transformations of the Poincar\'{e} group.

\section{Reference frames, spacetime splitting, and Electrodynamics}\label{sec: reference frames and electrodynamics}

Here, we will focus our attention on the role of reference frames in dealing with classical relativistic Electrodynamics. 
Usually, when a reference frame on the spacetime manifold is fixed, it is possible to recover the ``original'' formulation of Maxwell's equations in terms of $\mathrm{E}$, $\mathrm{B}$, $\mathrm{D}$ and $\mathrm{H}$ by means of a splitting procedure for the relativistic Maxwell's equations involving the Faraday form $\mathrm{F}$ and the Ampere form $\mathrm{G}$ (see, for instance, \cite{ParrotElectrodynamics, AuchmannKurz, Fecko, MarmoShanxi, HehlObukhov, DeFeliceBini}).

We will now briefly recall this procedure in the case of the covariant formulation of Electrodynamics on Minkowski spacetime $(\mathcal{M},\,g)\cong(\mathbb{R}^4,\,\eta)$ to set the stage for the proof of our main result given in the following section.
As remarked in the introduction, the contents in this section are certainly not new and the aim of this section is purely pedagogical in spirit.

We start with the Faraday  $2$-form $\mathrm{F}$ satisfying the homogeneous equation 
\be \label{Eq: Maxwell F}
\mathrm{dF}=0.
\ee
Then, once the inertial reference frame $\mathcal{R}_{0}$ of example \ref{ex: inertial reference frame on minkowski spacetime} is selected, a decomposition analogous to that given in equation \eqref{Eq:DecompositionDifferentialOfDifferentialForms} can be performed. The equation $\mathrm{dF} = 0$ is then equivalent to the two equations
\be \label{Eq:MaxwellSplitted}
\begin{cases}
(\mathrm{dF})_{\perp} = \mathrm{d}_\perp \mathrm{F}_\perp \equiv \mathrm{d}_\perp \mathrm{B} =  0 \\
(\mathrm{dF})_{\parallel} = \mathcal{L}_{\Gamma_{0}} \mathrm{F}_{\perp} - \mathrm{d}_{\perp} i_{\Gamma_{0}} \mathrm{F} \equiv \mathcal{L}_{\Gamma_{0}} \mathrm{B} - \mathrm{d}_{\perp} \mathrm{E} = 0
\end{cases}
\ee
where $\mathrm{E}$ and $\mathrm{B}$ are a $1-$form and a $2-$form over the spacetime that have the following local expression
\be\label{Eq:MaxwellSplitted2}
\begin{array}{l} \mathrm{E}\,:=\,i_{\Gamma_{0}}\mathrm{F}\,=\, - E_{1}dx^{1} - E_{2}dx^{2} - E_{3}dx^{3} \\ \\ \mathrm{B}\,:=\, \mathrm{F}_{\perp}\,=\, B_{3}dx^{1}\wedge dx^{2} - B_{2}dx^{1}\wedge dx^{3} + B_{1}dx^{2}\wedge dx^{3} \end{array}
\ee
Then, equations \eqref{Eq:MaxwellSplitted} split into
\be
0=\mathrm{d}_\perp \mathrm{B}=(\partial_{1}B_{1} + \partial_{2}B_{2} + \partial_{3}B_{3})dx^{1}\wedge dx^{2} \wedge dx^{3}\,,
\ee
\be
\begin{split}
0=\mathcal{L}_{\Gamma_{0}} \mathrm{B} - \mathrm{d}_{\perp} \mathrm{E}\,=  &(\partial_{2}E_{3} - \partial_{3}E_{2} + \partial_{0}B_{1})dx^{2}\wedge dx^{3}+ \\
+ &(\partial_{3}E_{1} - \partial_{1}E_{3} + \partial_{0}B_{2})dx^{3}\wedge dx^{1}+ \\ 
+ &( \partial_{1}E_{2} - \partial_{2}E_{1} + \partial_{0}B_{3})dx^{1}\wedge dx^{2}\,,
\end{split}
\ee
which are essentially the first two Maxwell equations in the original formulation of equation \eqref{eqn: original Maxwell}, i.e., the \textit{absence of magnetic monopoles} and the \textit{Faraday-Lenz  equations}. 
A similar procedure may be carried on starting from the equation 
\be \label{Eq: Maxwell G}
\mathrm{dG} = \mathrm{J}
\ee
for the Ampere $2$-form $\mathrm{G}$, obtaining the other two Maxwell equations involving the fields $D$ and $H$, i.e., the \textit{Ampére law} (including the displacement term added by Maxwell) and the \textit{Gauss' law}.
Clearly, when Minkowski spacetime is replaced by a ``more general'' spacetime, we no longer have the inertial reference frame $\mathcal{R}_{0}$ and all previous coordinate expressions are valid only locally in a coordinate system adapted to the given reference frame.
However, we may always define the fields $\mathrm{E},\,\mathrm{B},\, \mathrm{D},\, \mathrm{H}$ because of the intrinsic character of the language of differential forms.

At this point, it is worth stressing that what are usually referred to be the \textit{Maxwell equations} are the equations \eqref{Eq: Maxwell F} and \eqref{Eq: Maxwell G} together with a set of constitutive relations between $\mathrm{F}$ and $\mathrm{G}$\footnote{For a more detailed discussion on the constitutive relations see, for instance, \cite{ParrotElectrodynamics, Scheck, Deschamps_Electromagnetics_and_differential_forms} .}. In the next, we will always use the constitutive relations of the vacuum, namely, the ones induced by the \textit{Hodge}-$\star$ \textit{operator} associated to the \textit{Minkowski metric}
\be 
\mathrm{G}\,=\, \star \mathrm{F} \,.
\ee
The choice of these specific kind of constitutive relations is not too restrictive, it is justified because we aim to deal with Electromagnetic theory at a fundamental level.
In any case, those constitutive equations which may be described by a Hodge-type operator  are ``local'' constitutive equations, for instance, they exclude ``hysteresis'' phenomena.

\vsp

In the next section we would like to understand if it is possible to obtain a full-fledged covariant formulation of Maxwell's theory of Electrodynamics if we consider only the constraint equations, but in different reference frames. To the best of our knowledge this issue has not been addressed in the literature and the answer we will provide is in the affirmative. In other words, we will show that from the knowledge of the Gauss-law and the conservation of magnetic flux in suitably chosen reference frames, one can reconstruct the information contained in the dynamical equations and the result is the covariant formulation which we recalled in this section.

\section{From constraints to the covariant formulation of Electrodynamics} \label{Sec: from constraints to covariant}

In this section, we want to show how to obtain the covariant formulation of \textit{Classical Electrodynamics} in terms of the \textit{Faraday $2-$form} $\mathrm{F}$, the \textit{Ampere $2-$ form}  $\mathrm{G}$ and the $3-$form $\mathrm{J}$ called \textit{four-current}, by means of two constraint equations involving the electric induction, the magnetic induction and the charge density as perceived by a sufficient number of observers.
Specifically, we will address this problem in the context of \textit{Classical Electrodynamics} formulated on Minkowski spacetime $(\mathcal{M},\,g)\cong(\mathbb{R}^4,\,\eta)$.

Before proceeding further, it is worth stressing the fact that the procedure of ``covariantization'' we are about to propose is coherent only if it is intended at the microscopical level. 
Specifically, the only kind of sources considered in our procedure are those charges  described by means of $3$-forms over the spatial leaf associated with each observer. We also assume that there exists a fiducial reference frame where the considered distribution is at rest.
This is equivalent to say that only an electric induction is perceived by the observer associated with this fiducial reference frame. 
Referring to the covariantization procedure explained below, these kinds of sources are the ones that are the pull-back (by means of the immersion of the spatial leaves associated with the observers) of a time-like current on $\mathcal{M}$. Consequently, the only kind of currents we admit are the so called \textit{convection currents}, namely, the currents generated by observing the charge distribution from a moving frame (with respect to the fiducial one) where elementary charges appear in motion and generate a magnetic field. 
At the level of the fields, this means that we admit only two possibilities, the first one is realized in the fiducial frame where only the electric induction exists, while the second one is realized in all the other inertial frames where the previous electric induction is seen as a combination of an electric induction and a magnetic field produced by the convection current $\mathrm{j}$. 
At the level of the invariants of the Electromagnetic field (see \cite[pp. 92/94]{LandauCampi}  and \cite[p. 163]{Scheck}), this means that we only deal with situations where the first invariant associated with $\mathrm{G}$ is
\be
\mathrm{G} \wedge \mathrm{G} \,=\,  \left( \, \textbf{D}\cdot \textbf{H} \, \right)  \mathrm{d}^4\textbf{x}\,=\,0 ,
\ee
while the second  invariant is 
\be
\mathrm{G} \wedge \star \mathrm{G} \,= \, \left( \,|\,\textbf{D}\,|^2 \,- \, |\,\textbf{H}\,|^2 \,\right) \mathrm{d}^4\textbf{x}\,\geq 0 .
\ee 
The limit situation where the two invariants are both zero coincides with the absence of sources. 
In this case we would deal with pure radiation fields.

On the other hand, the covariantization procedure we will describe does not impose, in principle, any constraint on the values of the other invariants of the Electromagnetic field containing $\mathrm{F}$
\be
\mathrm{F}\wedge \mathrm{F} \qquad \mathrm{F} \wedge \star \mathrm{F} \qquad \mathrm{F} \wedge \mathrm{G} \,,
\ee
because $\mathrm{F}$ is involved in the homogeneous equation where sources do not appear. However,  $\mathrm{F}$ and $\mathrm{G}$ are not actually independent. Indeed, they are related by means of a set of constitutive relations. Specifically, since we are dealing with Electromagnetic theory at a fundamental level, we assume the constitutive relations to be the ones in the vacuum described at the end of sections \ref{sec: reference frames and electrodynamics}. Thus, the limitations on the values of the invariants associated with $\mathrm{G}$ reflect into limitations on the values of the invariants associated with $\mathrm{F}$. Actually, the five invariants written above collapse into only two invariants when the constitutive relations are used.

The case we did not consider, namely, the case where the second invariant associated with $\mathrm{G}$ is negative, may be possible only if magnetic monopoles  are admitted, or in models where there is a current but no net charge. 
The latter situation, as it was proved in \cite{BohrPhDThesis, vanLeeuwen} by \textit{Bohr} and \textit{Van Leuween}, is possible only at the quantum level and, thus, can not be included in our classical formulation. 
In the same way, also particles with spin possessing an intrinsic magnetic moment (sources of magnetic fields) are not included in our discussion.

\vsp

Let us start with Minkowski spacetime $(\mathcal{M},\,g)\cong(\mathbb{R}^4,\,\eta)$ and let us fix the inertial reference frame $\mathcal{R}^{(0)}=\Gamma^{(0)}\otimes\theta^{(0)}$  built in example \ref{ex: inertial reference frame on minkowski spacetime}.
We will focus on the family of inertial reference frames on $\mathcal{M}$ that may be obtained acting with elements of the Poincar\'{e} group on the fiducial reference frame $\mathcal{R}^{(0)}$.
In particular, it will be sufficient to consider the subfamily 
$$
\{\mathcal{R}^{(\mu)}=\Gamma^{(\mu)}\otimes\theta^{(\mu)}\}_{\mu=0,1,2,3}
$$
of inertial reference frames such that $\mathcal{R}^{(1)}$ is obtained from $\mathcal{R}^{(0)}$ by means of a Lorentz boost along the $x^{(0)}_1$-direction, $\mathcal{R}^{(2)}$ is obtained from $\mathcal{R}^{(0)}$ by means of a Lorentz boost along the $x^{(0)}_2$-direction, and $\mathcal{R}^{(3)}$ is obtained from $\mathcal{R}^{(0)}$ by means of a Lorentz boost along the $x^{(0)}_3$-direction (see  example \ref{ex: inertial reference frame on minkowski spacetime}). 
Essentially, as we wrote in section \ref{Sec: reference frames}, given a reference frame $\mathcal{R}^{(\mu)}$, the tangent space at each point of $\mathcal{M}$ is spanned by the set 
$$
\{\,\Gamma^{(\mu)},\, X^{(\mu)}_1,\,X^{(\mu)}_2,\,X^{(\mu)}_3\,\}
$$
of vector fields, where the set $\{\,X^{(\mu)}_j\,\}_{j=1,2,3}$ is a basis of the kernel of $\theta^{(\mu)}$ as a module. 
The representation of a Poincaré boost along the $x^{(0)}_j$ direction on $\mathcal{M}$ is the following
\be
\begin{split}
t^{(j)} &\,=\, \gamma \, t^{(0)} - \beta \gamma \, x^{(0)}_j \\
x^{(j)}_j &\,=\, -\beta \gamma \, t^{(0)} + \gamma \, x^{(0)}_j \\
x^{(j)}_k &\,=\, x^{(0)}_k \;\; \text{for } \, k \neq j ,
\end{split}
\ee 
while its inverse is
\be
\begin{split}
t^{(0)} &\,=\, \gamma \, t^{(j)} + \beta \gamma \, x^{(j)}_j \\
x^{(0)}_j &\,=\, \beta \gamma \, t^{(j)} + \gamma \, x^{(j)}_j \\
x^{(0)}_k &\,=\, x^{(j)}_k \;\; \text{for } \, k \neq j \,.
\end{split}
\ee
What we are assuming is that
\be \label{Eq: boost fields}
\begin{split}
\Gamma^{(j)} \,=\, \gamma \, \Gamma^{(0)} + \beta \gamma \, X^{(0)}_j \\
X^{(j)}_j \,=\, \beta \gamma \, \Gamma^{(0)} + \gamma \, X^{(0)}_j \\
X^{(j)}_k \,=\, X^{(0)}_k  \;\; \text{for } \, k \neq j \,.
\end{split}
\ee
Each of these four reference frames gives a different splitting of the spacetime manifold together with the decomposition 
\be
\mathcal{M}\,=\,\mathcal{T}^{(\mu)}\,\times\,\mathcal{S}^{(\mu)}
\ee
of $\mathcal{M}$ in terms of time and space manifolds. 
We denote with $i^{(\mu)}_{t_{\mu}}$ the canonical immersion of the space manifold $\mathcal{S}^{(\mu)}$ in $\mathcal{M}$ at the time instant $t_{\mu}\in\mathcal{T}^{(\mu)}$ and with $\mathrm{d}^{(\mu)}_\perp$ the derivation we defined in the end of section \ref{Sec: reference frames} associated with the foliation of the reference frame $\mathcal{R}^{(\mu)}$.

\vsp

Let $\mathrm{F}$ (usually called \textit{Faraday form}) be a 2-form  on the spacetime describing the electromagnetic field, and let us assume that a family of observers are measuring it. 
The covariant version of Maxwell equations for $\mathrm{F}$ is 
\be
\mathrm{dF}=0 ,
\ee
which is an equation involving a 3-form. 
In the following, we will prove that if one chooses the family of reference frames just defined, it is possible to reconstruct the whole set of covariant equations once the transversal constraint 
\be
\mathrm{d}_{\perp}\mathrm{F}_{\perp}=0
\ee
is valid in any reference frame (namely the magnetic field "seen" from this frame). 

Roughly speaking, the knowledge of the transversal components of a differential form in different, suitably selected reference frames, allows to reconstruct the whole differential forms. 
However, the minimal number of reference frames required for the reconstruction depends on the degree of the form. 
We want to reconstruct a 3-form and in this case 4-different reference frames are needed. 

The main results of this section will be a consequence of the following proposition:
\begin{prop} \label{Thm: reconstruction of electric field from magnetic fields}
Consider a family 
$$
\{\mathcal{R}^{(\mu)}=\Gamma^{(\mu)}\otimes\theta^{(\mu)}\}_{\mu=0,1,2,3}
$$
of reference frames on Minkowski spacetime $\mathcal{M}$ obeying equation \eqref{Eq: boost fields}. 
Consider a differential 3-form $\mathrm{J}$ on $\mathcal{M}$ which may be decomposed as
\be
\mathrm{J} \,=\, \theta^{(\mu)} \wedge \mathrm{J}_\parallel^{(\mu)} + \mathrm{J}_\perp^{(\mu)} \,,
\ee
where $\mu$ is the index labelling the reference frame in the previously defined family with respect to which the decomposition is taken.
If the differential $3$-form $\mathrm{J}_\perp^{(\mu)}$ is known for all $\mu \,=\, 0,\,...,\,3$,  it is possible to reconstruct the differential 3-form $\mathrm{J}$.

\begin{proof}
Let us start noting that 
\be
\mathrm{J}^{(1)}_\perp \left(\, X^{(1)}_1,\,X^{(1)}_2,\,X^{(1)}_3 \, \right) \,=\, \mathrm{J}\left(\, X^{(1)}_1,\,X^{(1)}_2,\,X^{(1)}_3 \, \right) \,
\ee
because $ X^{(1)}_j$ is in the kernel of $\theta^{(1)}$ for every $j=1,2,3$.
Moreover, thanks to equation \eqref{Eq: boost fields}, we have
\be
X^{(1)}_1 \,=\, \beta \gamma \, \Gamma^{(0)} + \gamma \, X^{(0)}_1 
\ee
and 
\be
X^{(1)}_2 \,=\, X^{(0)}_2 \,, \qquad X^{(1)}_3\, = \, X^{(0)}_3\,.
\ee
Consequently, we may write
\begin{eqnarray}
&\mathrm{J}^{(1)}_\perp \left(\, X^{(1)}_1,\,X^{(1)}_2,\,X^{(1)}_3 \, \right) \,=\, \mathrm{J}\left( \,\beta \gamma \, \Gamma^{(0)} + \gamma \, X^{(0)}_1,\, X^{(0)}_2,\,X^{(1)}_3 \, \right) \,=\, \nonumber\\
& =\,\beta \gamma \, \mathrm{J}\left( \, \Gamma^{(0)},\, X^{(0)}_2,\,X^{(0)}_3 \, \right) + \gamma \, \mathrm{J}\left( \,X^{(0)}_1,\,X^{(0)}_2,\,X^{(0)}_3 \, \right) \,= \nonumber \\
& =\,\beta \gamma \, i_{\Gamma^{(0)}}\mathrm{J}\left(\,X^{(0)}_2,\,X^{(0)}_3 \, \right) + \gamma \, \mathrm{J}^{(0)}_\perp \left(\, X^{(0)}_1,\,X^{(0)}_2,\,X^{(0)}_3 \, \right) \,=\, \nonumber \\
& =\, \beta \gamma \, \mathrm{J}^{(0)}_\parallel \left(\, X^{(0)}_2,\,X^{(0)}_3 \, \right) + \gamma \, \mathrm{J}^{(0)}_\perp \left( \,X^{(0)}_1,\,X^{(0)}_2,\,X^{(0)}_3 \, \right) \,,
\end{eqnarray}
which is equivalent to
\be 
\begin{split}
\mathrm{J}^{(0)}_\parallel \left(X^{(0)}_2,\,X^{(0)}_3 \right) \, =\, \frac{1}{\beta} \left[\, \frac{1}{\gamma} \right.&\mathrm{J}^{(1)}_\perp \left(\, X^{(1)}_1,\,X^{(1)}_2,\,X^{(1)}_3 \, \right)  + \\
- & \left. \mathrm{J}^{(0)}_\perp \left( \, X^{(0)}_1,\,X^{(0)}_2,\,X^{(0)}_3 \, \right) \,\right] .
\end{split}
\ee
This expression only depends on the known quantities $\mathrm{J}^{(1)}_\perp \left( \, X^{(1)}_1,\,X^{(1)}_2,\,X^{(1)}_3 \, \right)$ and $\mathrm{J}^{(0)}_\perp \left(\, X^{(0)}_1,\,X^{(0)}_2,\,X^{(0)}_3 \, \right)$.

With the same algorithm, we can obtain the other components of $\mathrm{J}^{(0)}_\parallel $ obtaining
\be
\begin{split}
 \mathrm{J}^{(0)}_\parallel \left(\, X^{(0)}_1,\,X^{(0)}_2 \, \right)   &\,=\,\frac{1}{\beta} \left[\,\frac{1}{\gamma} \mathrm{J}^{(3)}_\perp \left(\, X^{(3)}_1,\,X^{(3)}_2,\,X^{(3)}_3 \, \right)+ \right. \\
 &\left. - \mathrm{J}^{(0)}_\perp \left(\, X^{(0)}_1,\,X^{(0)}_2,\,X^{(0)}_3 \, \right) \,\right]  \\
\mathrm{J}^{(0)}_\parallel \left(\, X^{(0)}_1,\,X^{(0)}_3 \, \right) &\,=\,\frac{1}{\beta} \left[\,-\frac{1}{\gamma} \mathrm{J}^{(2)}_\perp \left(\, X^{(2)}_1,\,X^{(2)}_2,\,X^{(2)}_3 \, \right) + \right. \\ 
&\left. +\mathrm{J}^{(0)}_\perp \left(\, X^{(0)}_1,\,X^{(0)}_2,\,X^{(0)}_3 \, \right) \,\right] \\
\mathrm{J}^{(0)}_\parallel \left(\, X^{(0)}_2,\,X^{(0)}_3 \, \right) &\,=\,\frac{1}{\beta} \left[\,\frac{1}{\gamma} \mathrm{J}^{(1)}_\perp \left(\, X^{(1)}_1,\,X^{(1)}_2,\,X^{(1)}_3 \, \right) +\right. \\
&\left. -\mathrm{J}^{(0)}_\perp \left(\, X^{(0)}_1,\,X^{(0)}_2,\,X^{(0)}_3 \, \right) \,\right] \,.
\end{split}
\ee
Consequenlty, we obtained all the elements needed to decompose $\mathrm{J}$ in the reference frame $\mathcal{R}^{(0)}$ only in terms of the transversal components in all the reference frames  $\mathcal{R}^{(\mu)}$ with $\mu=0,...,3$. 
\end{proof}
\end{prop} 

With a similar procedure, one can reconstruct forms of different degrees. 
However, the minimal number of reference frames required for the procedure to work  will change since the number of indipendent components of the transversal form depends on the degree of the form. 
For a 3-form, we have only one transversal component and three temporal components, and thus  we needed four reference frames connected via Lorentzian boosts. 
A 2-form (such as the \textit{Faraday} or \textit{Ampere form}), for instance, could be reconstructed using only three different reference frames because in any of them we are able to recover two temporal components out of three. 

From the physical point of view, a measure of magnetic flux performed by an experimenter gives a physical quantity as perceived in a given splitting of spacetime associated with a given reference frame $\mathcal{R}$. 
Using these data, the theoretician ``builds'' a differential $2$-form, say $\mathrm{B}$, on the spatial leaves of the foliation induced by $\mathcal{R}$.
The coefficients of this 2-form will depend on time according to the duration of the experiment. 
From the mathematical point of view, this means that the magnetic field may be modelled as a differential $2$-form over the space manifold $\mathcal{S}$ whose coefficients depends also on $t$, that is, as an element of $\Omega^{2}_0(\mathcal{M})$, the $\mathcal{F}(\mathcal{M})$-bimodule of differential forms introduced in section \ref{Sec: reference frames}.
Consequently, the experimental situation leads to a reconstruction of the transversal component of the electromagnetic field in a given reference frame.
In this context,  proposition \ref{Thm: reconstruction of electric field from magnetic fields} implies that the knowledge of the magnetic field in a fiducial reference frame and in other $2$ particular ones is sufficient to reconstruct also the electric field in the fiducial reference frame. 
As we will see in the rest of this section, the same argument allows us to reconstruct dynamical equations starting from constraints as perceived in different reference frames.  

\begin{rem}
It is worth stressing the fact that, from the experimental point of view, one may only measure flux of magnetic fields for a finite amount of time. 
Therefore, by means of a real experiment, one is only constructing the differential form $\mathrm{F}^{(\mu)}_\perp$ defined on $I^{(\mu)}\times \mathcal{S}^{(\mu)}$, $I^{(\mu)}$ being a bounded interval of $\mathcal{T}^{(\mu)}$. A complete reconstruction over the whole $\mathcal{T}^{(\mu)}\times \mathcal{S}^{(\mu)}$ would require an idealized experiment performed for an infinite time, or some additional  regularity assumptions on the coefficients of $\mathrm{B}^{(\mu)}_{t_\mu}$ (such as the analicity). 
We do not want to indulge on this aspect anymore and we settle for a local reconstruction procedure.
\end{rem}

Let us now deal with the sources of the Electromagnetic field. 
A charge distribution is modelled as a differential $3$-form, say $\rho$, on the spatial leaves of the foliation associated with the reference frame of an observer.
This means that $\rho$ is an element of $\Omega_0^3(\mathcal{M})$. 
Measuring $\rho$ for $t \in I$ means to reconstruct the transversal component of the differential form $\mathrm{J}$ on $I \times \mathcal{S}$. 
Now, if we assume that an electromagnetic source, say $\mathrm{J}$, exists and that we have measured its transversal components in all the reference frames of proposition  \ref{Thm: reconstruction of electric field from magnetic fields}, it is possible to reconstruct the whole $\mathrm{J}$ in $\mathcal{R}$.
From the experimental point of view, this means that we are able to reconstruct the whole source $\mathrm{J}$ of the Electromagnetic field by measuring charge distributions in $4$ particular reference frames.  

\vsp

We are now ready to prove our main result, namely, we apply the previous procedure to constraint equations in order to obtain the full set of dynamical equations describing Classical Electrodynamics.   
Let us assume that an electromagnetic field, say $\mathrm{F}$, exists and that the \textit{broken magnet experiment} is performed in the four reference frames of proposition  \ref{Thm: reconstruction of electric field from magnetic fields}. 
This means that, introducing the quantity 
\be
\mathrm{B}^{(\mu)}_{t_{\mu}}:= \left( i^{(\mu)}_{t_{\mu}} \right)^* \mathrm{B}^{(\mu)} 
\ee
with $\mathrm{B}^{(\mu)}  = \mathrm{F}^{(\mu)}_\perp$, the flux
\be
\Phi_{\Sigma^{(\mu)}} \left( \mathrm{B}^{(\mu)} \right) \,:=\, \int_{\Sigma^{(\mu)}} \, \mathrm{B}^{(\mu)}_{t_{\mu}}  \,,
\ee
of $\mathrm{B}^{(\mu)}$ through a two-dimensional closed, boundaryless surface, say $\Sigma^{(\mu)}$, in $\mathcal{S}^{(\mu)}$ is measured and is seen to be $0$ for all $\Sigma^{(\mu)}$. 
By means of Stokes' theorem we have
\be \label{Eq: constraints magnetic field}
\mathrm{d}\mathrm{B}^{(\mu)}_{t_{\mu}} \,=\, 0 \,,
\ee
where $\mathrm{d}$ is the differential on $\Sigma^{(\mu)}$ and $t_{\mu}\in I^{(\mu)}$. 

These are the constraint equations for the magnetic field in the four reference frames considered. Since equation \eqref{Eq: constraints magnetic field} is valid for all $t_\mu \in I^{(\mu)}$ we can replace equation \eqref{Eq: constraints magnetic field} with the equation
\be
\mathrm{d}^{(\mu)}_{\perp}F^{(\mu)}_{\perp}\,=\,\mathrm{d}^{(\mu)}_{\perp}B^{(\mu)} = 0\,
\ee
because, by definition, we have
\be
\mathrm{d} \mathrm{B}^{(\mu)}_{t_{\mu}} = \left( i^{(\mu)}_{t_{\mu}} \right)^* \left( \mathrm{d}^{(\mu)}_{\perp} B^{(\mu)}\right)\;\;\forall t_{\mu}\in I^{(\mu)} ,
\ee
and the only indipendent component of $\mathrm{d}^{(\mu)}_{\perp}B^{(\mu)}$ is $\mathrm{d}_{\perp}^{(\mu)}B^{(\mu)}(\,X^{(\mu)}_1,\, X^{(\mu)}_2,\,X^{(\mu)}_3\,)$. Consequently, applying proposition \ref{Thm: reconstruction of electric field from magnetic fields} to the differential form $\mathrm{dF}$, we have that  the constraint equations 
\be
(\,\mathrm{dF}\,)^{(\mu)}_\perp \,=\, \mathrm{d}^{(\mu)}_\perp \mathrm{F}^{(\mu)}_\perp \,=\,\mathrm{d}^{(\mu)}_{\perp}B^{(\mu)}\,=\,0
\ee
valid in all the four reference frames are equivalent to the full set of dynamical equations 
\be
\begin{cases}
(\,\mathrm{dF}\,)^{(0)}_\parallel \,=\, 0 \\
(\,\mathrm{dF}\,)^{(0)}_\perp \,=\, 0
\end{cases}
\implies \;\; \mathrm{dF} \,=\,0
\ee
in the reference frame $\mathcal{R}^{(0)}$.
From the physical point of view, this means that a suitable "covariantization" of the constraint equation regarding the magnetic field is enough to recover the full set of homogeneous Maxwell's equations. 

\vsp

Concerning inhomogeneous Maxwell equations a similar construction may be performed. We assume that an electromagnetic field and a source exist and that we are measuring them from the different reference frames of proposition \ref{Thm: reconstruction of electric field from magnetic fields}. 
Now, analogously to the magnetic field, we set
\be
\begin{split}
\mathrm{D}^{(\mu)}_{t_{\mu}} &:= \left( i^{(\mu)}_{t_{\mu}} \right)^* D^{(\mu)} \\
\rho^{(\mu)}_{t_{\mu}} &:= \left( i^{(\mu)}_{t_{\mu}} \right)^* \rho^{(\mu)},
\end{split}
\ee
and we assume that \textit{Gauss' law} is observed in each reference frame.
Equivalently, we assume the validity of 
\be 
\int_{\partial \Omega^{(\mu)}} \mathrm{D}^{(\mu)}_{t_\mu} \,=\, \int_{\Omega^{(\mu)}} \mathrm{\rho}^{(\mu)}_{t_\mu}
\ee    
for every volume, say $\Omega^{(\mu)}$, in $\mathcal{S}^{(\mu)}$.
The local version of this equation reads 
\be
\mathrm{dD}^{(\mu)}_{t_\mu} \,=\, \mathrm{\rho}^{(\mu)}_{t_\mu}
\ee
for any reference frame in the family considered. 
Repeating the same arguments used above, we obtain the following equations
\be
(\, \mathrm{dG} \, )^{(\mu)}_\perp \,=\, (\,\mathrm{J}\,)^{(\mu)}_\perp \,,
\ee
where $\mathrm{G}$ is the differential $2$-form (usually called \textit{Ampere form}) on the spacetime associated with $\mathrm{D}^{(\mu)}$ while $\mathrm{J}$ is the differential $3$-form on the spacetime associated with $\mathrm{\rho}^{(\mu)}$. 
Then, proposition \ref{Thm: reconstruction of electric field from magnetic fields}  implies that
\be
\begin{split}
(\,\mathrm{dG}\,)^{(0)}_\parallel \left(\, X^{(0)}_1,\,X^{(0)}_2 \, \right) \,=\,&\frac{1}{\beta} \left[\,\frac{1}{\gamma} \mathrm{d}^{(3)}_\perp \mathrm{G}^{(3)}_\perp \left(\, X^{(3)}_1,\,X^{(3)}_2,\,X^{(3)}_3 \, \right) \right. \\ 
& \left. - \mathrm{d}^{(0)}_\perp \mathrm{G}^{(0)}_\perp \left(\, X^{(0)}_1,\,X^{(0)}_2,\,X^{(0)}_3 \, \right) \,\right] \,=\, \\
\,=\,& \frac{1}{\beta} \left[\,\frac{1}{\gamma} \mathrm{J}^{(3)}_\perp \left(\, X^{(3)}_1,\,X^{(3)}_2,\,X^{(3)}_3 \, \right) \right. \\
&\left. - \mathrm{J}^{(0)}_\perp \left(\, X^{(0)}_1,\,X^{(0)}_2,\,X^{(0)}_3 \, \right) \,\right] \,=\, \\
\,=\,& \mathrm{J}^{(0)}_\parallel \left(\,X^{(0)}_1,\,X^{(0)}_2 \,\right)  \\ & \\
(\,\mathrm{dG}\,)^{(0)}_\parallel \left(\, X^{(0)}_1,\,X^{(0)}_2 \, \right) \,=\,& \frac{1}{\beta} \left[\,-\frac{1}{\gamma} \mathrm{d}^{(2)}_\perp \mathrm{G}^{(2)}_\perp \left(\, X^{(2)}_1,\,X^{(2)}_2,\,X^{(2)}_3 \, \right) \right. \\
&\left. + \mathrm{d}^{(0)}_\perp \mathrm{G}^{(0)}_\perp \left(\, X^{(0)}_1,\,X^{(0)}_2,\,X^{(0)}_3 \, \right) \,\right] \,=\, \\
\,=\, & \frac{1}{\beta} \left[\,-\frac{1}{\gamma} \mathrm{J}^{(2)}_\perp \left(\, X^{(2)}_1,\,X^{(2)}_2,\,X^{(2)}_3 \, \right) \right. \\
& \left. + \mathrm{J}^{(0)}_\perp \left(\, X^{(0)}_1,\,X^{(0)}_2,\,X^{(0)}_3 \, \right) \,\right] \,=\, \\
 \,=\,&  \mathrm{J}^{(0)}_\parallel \left(\,X^{(0)}_1,\,X^{(0)}_3 \,\right) \\ & \\
(\,\mathrm{dG}\,)^{(0)}_\parallel \left(\, X^{(0)}_1,\,X^{(0)}_2 \, \right) \,=\,& \frac{1}{\beta} \left[\,\frac{1}{\gamma} \mathrm{d}^{(1)}_\perp \mathrm{G}^{(1)}_\perp \left(\, X^{(1)}_1,\,X^{(3)}_2,\,X^{(1)}_3 \, \right) \right. \\
&\left. - \mathrm{d}^{(0)}_\perp \mathrm{G}^{(0)}_\perp \left(\, X^{(0)}_1,\,X^{(0)}_2,\,X^{(0)}_3 \, \right) \,\right] \,=\, \\
\,=\, & \frac{1}{\beta} \left[\,\frac{1}{\gamma} \mathrm{J}^{(1)}_\perp \left(\, X^{(1)}_1,\,X^{(1)}_2,\,X^{(1)}_3 \, \right) \right. \\
&\left. - \mathrm{J}^{(0)}_\perp \left(\, X^{(0)}_1,\,X^{(0)}_2,\,X^{(0)}_3 \, \right) \,\right] \,=\, \\
 \,=\,& \mathrm{J}^{(0)}_\parallel \left(\,X^{(0)}_2,\,X^{(0)}_3 \,\right) \,,
\end{split}
\ee
and these equations together with $(\, \mathrm{dG} \, )^{(0)}_\perp \,=\, (\,\mathrm{J}\,)^{(0)}_\perp$ are equivalent to
\be
\begin{cases}
\mathrm{dG}^{(0)}_\parallel \,=\, \mathrm{J}^{(0)}_\parallel \\
\mathrm{dG}^{(0)}_\perp \,=\, \mathrm{J}^{(0)}_\perp
\end{cases}
\implies \;\; \mathrm{dG} \,=\, \mathrm{J}
\ee
Again, in a suggestive way, we may conclude that a suitable "covariantization" of the Gauss' law leads to the inhomogeneous Maxwell's equations.

\section{Conclusions}

We proposed a covariantization procedure by means of which it is possible to recover the spacetime formulation of Maxwell's equations on Minkowski spacetime   from the  local version of the law of conservation of magnetic flux and the Gauss law as ``perceived'' by the observers associated with different inertial reference frames on Minkowski spacetime.
Even if these laws do not involve time derivatives, and thus may be interpreted as constraints on the Cauchy data in a given inertial reference frame, it is possible to extract ``dynamical'' information from them as soon as we consider constraint equations with respect to different inertial reference frames.
In particular, if one looks at the two constraint equations from all possible inertial reference frames on Minkwoski spacetime, one is able to obtain the other two equations, i.e., Faraday-Neumann and Ampere equations. 
In a nutshell, the spacetime formulation of Maxwell's equations on Minkowski spacetime is obtained as the result of a ``covariantization'' procedure for the Gauss' law and the local version of the law of conservation of magnetic flux.

We want to stress the fact that, in the framework of Lagrangian theories described by  a degenerate Lagrangian, the dynamics is invariant under some gauge group of transformations and the equations of motion split into genuine evolutionary equations and constraints. 
With this in mind, the next step to take is to apply the covariantization procedure outlined in this paper to the case of other, not-necessarily  Abelian, gauge theories.
We plan to address this instance in future works.

\bigskip

\noindent{\bf Acknowledgements} 
G.M. would like to thank the support provided by the Santander/UC3M Excellence Chair Programme 2019/2020; he also  acknowledges financial support from the Spanish Ministry of Economy and Competitiveness, through the Severo Ochoa Programme for Centres of Excellence in RD(SEV-2015/0554>. G.M. is a member of the Gruppo Nazionale di Fisica Matematica (INDAM), Italy. F.D.C. would like to thank partial support provided by the MINECO research project MTM2017-84098-P and QUITEMAD++, S2018/TCS-A4342.

\end{document}